\documentclass{emulateapj}
\usepackage{natbib}

\def\gsim{\;\rlap{\lower 2.5pt
 \hbox{$\sim$}}\raise 1.5pt\hbox{$>$}\;}
\def\lsim{\;\rlap{\lower 2.5pt
   \hbox{$\sim$}}\raise 1.5pt\hbox{$<$}\;}

\def\lengthunits{$h^{-1}\, \textrm{kpc}$ }
\def\massunits{$h^{-1}\, M_\odot$ }
\def\massunitsend{$h^{-1}\, M_\odot$}

\begin{document}

\title{On the Origin of Dynamically Cold Rings Around the Milky Way}
\shorttitle{Rings Around the Milky Way}
\author{Joshua D. Younger,\altaffilmark{1,2} Gurtina Besla,\altaffilmark{1} T.~J. Cox,\altaffilmark{1,3} Lars Hernquist,\altaffilmark{1} Brant Robertson,\altaffilmark{4,5,6} \& Beth Willman\altaffilmark{1,7}}
\altaffiltext{1}{Harvard--Smithsonian Center for Astrophysics, 60 Garden Street, 
Cambridge, MA 02138}
\altaffiltext{2}{jyounger@cfa.harvard.edu}
\altaffiltext{3}{Keck Foundation Fellow}
\altaffiltext{4}{Kavli Institute for Cosmological Physics and Department of Astronomy and Astrophysics, University of Chicago, 933 East 56th Street, Chicago, IL 60637}
\altaffiltext{5}{Enrico Fermi Insitute, 5640 South Ellis Avenue, Chicago, IL 60637}
\altaffiltext{6}{Spitzer Fellow}
\altaffiltext{7}{Clay Fellow}

\begin{abstract}

We present a scenario for the production of dynamically cold rings
around the Milky Way via a high--eccentricity, flyby encounter.  These
initial conditions are more cosmologically motivated than those
considered in previous works.  We find that the encounters we examine
generically produce a series of nearly dynamically cold ring--like
features on low--eccentricity orbits that persist over timescales of
$\sim 2-4$ Gyr via the tidal response of the primary galaxy to the
close passage of the satellite.  Moreover, they are both qualitatively
and quantitatively similar to the distribution, kinematics, and
stellar population of the Monoceros ring.  Therefore, we find that a
high eccentricity flyby by a satellite galaxy represents a
cosmologically appealing scenario for forming kinematically distinct
ring--like features around the Milky Way.

\end{abstract}

\keywords{galaxy: kinematics and dynamics, galaxy: structure, galaxies: interactions, galaxies: structure, methods: $n$--body simulations}

\section{Introduction}
\label{sec:intro}

\begin{figure*}
\plotone{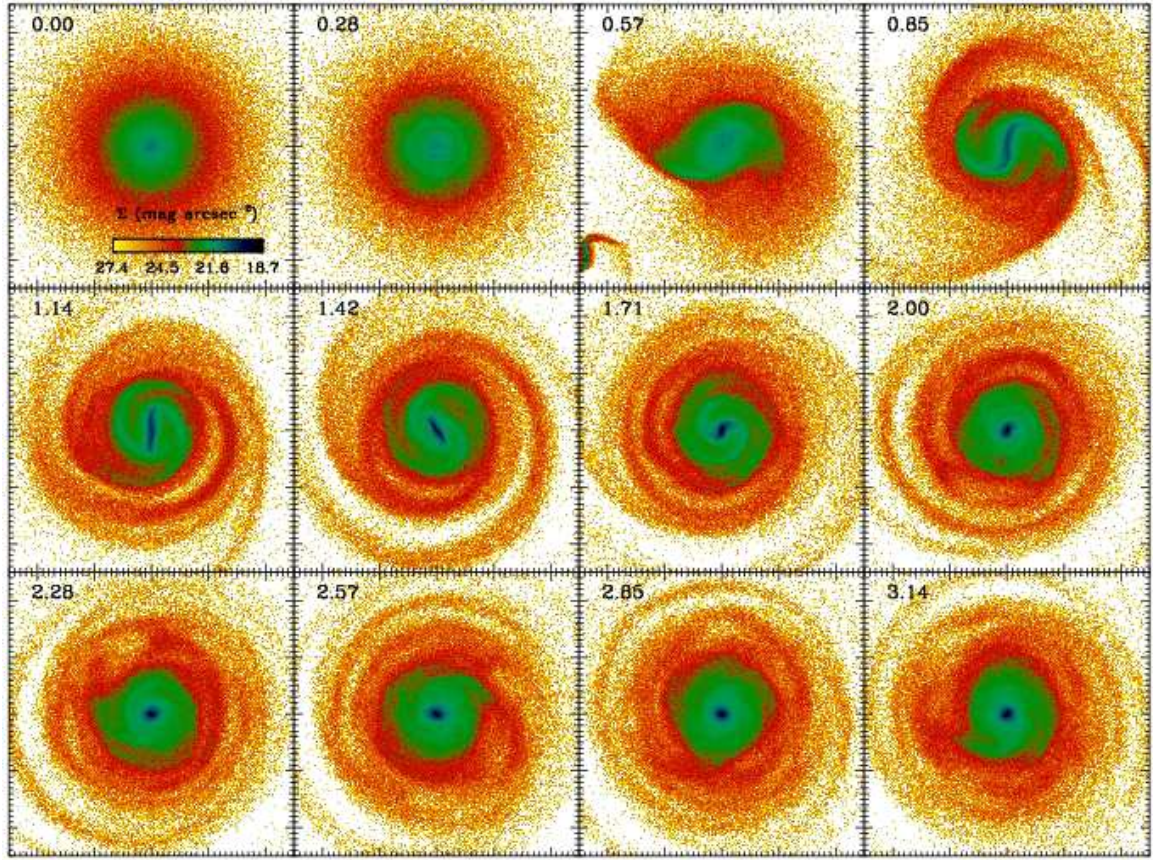}
\caption{The evolution of the projected stellar mass density, colored according to a logarithmic scale.  The panels are 70 kpc on a side, and the simulation time is printed in the upper left hand corner in units of Gyr.  The color-bar indicates the projected K--band surface brightness in units of mag arcsec$^{-2}$, assuming a constant $M/L$ ratio of 2 in solar units.}
\label{fig:allstars}
\end{figure*}

We live in a hierarchical universe, in which mergers are a frequent occurrence
\citep{laceycole1993,somerville1999,somerville2000}.  In this
scenario, galaxies like the Milky Way (MW) build up  
much of their mass by accreting smaller satellite galaxies.
These minor mergers are connected to a variety of
observable stellar structural and kinematic phenomena 
in the MW and external galaxies, including: the build--up of stellar
halos \citep[e.g.][]{bullock2005,bell2007}, so--called
``antitruncated" \citep[][]{erwin2005,pohlen2006,younger2007} and
``extended" outer disks \citep[][]{ibata2005,penarrubia2006,ibata2007}
, and the dynamical heating of the stellar disk
\citep[e.g.,][]{toth1992,quinn1993,walker1996,kazantzidis2007}.

In our own galaxy, \citet{newberg2002} recently identified a coherent
ring--like structure in Monoceros (MRi), using data from the Sloan
Digital Sky Survey \citep[SDSS: ][]{york2000}.  Since this initial
detection, the MRi has been identified in the infrared from the Two
Micron All--Sky Survey \citep[2MASS: ][]{strutskie2006} by
\citet{rochapinto2003} and \citet{martin2004a}, reanalyzed using
updated SDSS data \citep{grillmair2006,belokurov2006}, and has been
followed--up photometrically by several optical surveys
\citep{ibata2003,conn2005a,conn2007}.  The MRi subtends 
$\gsim 100^\circ$ in galactic longitude and lies at a
galactocentric radial distance of between 15 and 20 kpc.
Spectroscopic studies have shown that it is kinematically
distinct from the disk -- dynamically cold with a low
eccentricity orbit -- and is composed primarily of low--metallicity
stars with $-1.6 \lsim [{\rm Fe/H}] \lsim -0.4$
\citep{crane2003,yanny2003,conn2005b,martin2005,martin2006}.

Numerical modeling has suggested that similar structures could
be formed via tidal disruption of a low--mass satellite on a nearly
coplanar orbit
\citep{helmi2003,meza2005,martin2005,penarrubia2005,penarrubia2007}.
The most successful of these models argue for a very low mass
companion with $M_s/M_{MW} \sim 1/1000$ on a nearly circular ($e=0.1$)
prograde orbit \citep{martin2005,penarrubia2005,penarrubia2007}.
However, recent cosmological simulations \citep{benson2005,khochfar2006} have found that the
orbits of satellite galaxies are likely be highly eccentric.
The dynamical friction timescale for such interactions is very long
 and as a result it is unlikely that such
a satellite could have circularized in time to form the MRi
\citep{besla2007,boylankolchin2007}.  Therefore, while they are
successful at reproducing many of the characteristics of the MRi,
these models may have cosmologically unappealing initial conditions.

With this in mind, we propose an alternative mechanism for forming dynamically cold ring--like structures around the MW: a flyby encounter with a small satellite on a high eccentricity orbit.  This scenario was first suggested by \citet{kazantzidis2007}, but here we investigate the triggering event in detail, and discuss its relevance to the MRi.

\section{Simulations}
\label{sec:sims}

The simulations presented in this study were performed with {\sc Gadget2} \citep{springel2005}, an N--Body/SPH (Smooth Particle Hydrodynamics) code using the entropy conserving formalism of \citet{springelhernquist2002}.  We include radiative cooling and star formation, tuned to fit the observed local Schmidt Law \citep{schmidt1959,kennicutt1998}.  We also incorporate a sub--resolution multi--phase model of the interstellar medium (ISM) \citep{springelhernquist2003} -- softened ($q_{EoS} = 0.25$) such that the mass--weighted ISM temperature is $\sim 10^{4.5}$ K -- and sink particles representing supermassive black holes that accrete gas and release isotropic thermal feedback to self--regulate their growth \citep{springeldimatteo2005a}.  

\begin{figure*}
\plottwo{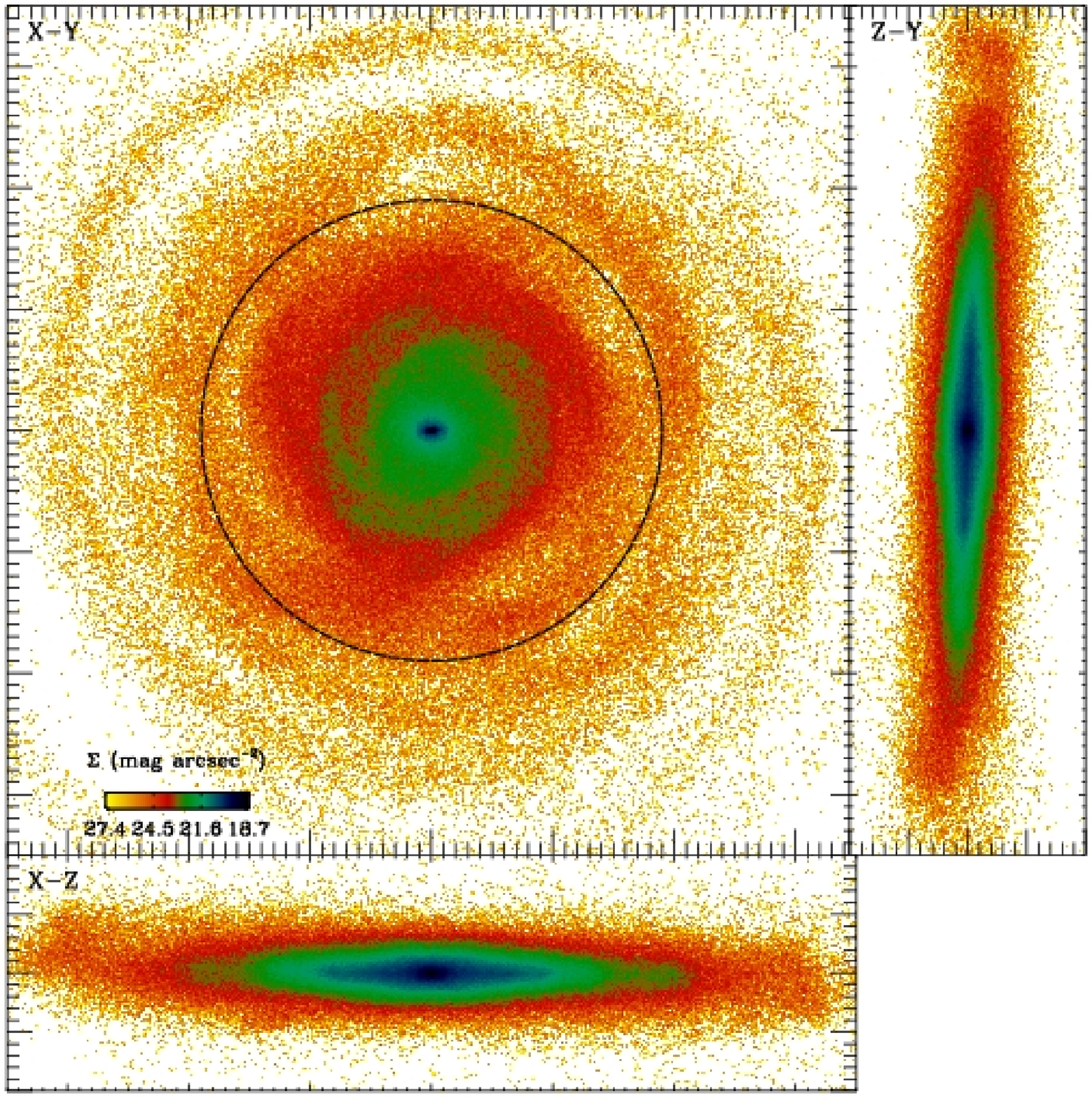}{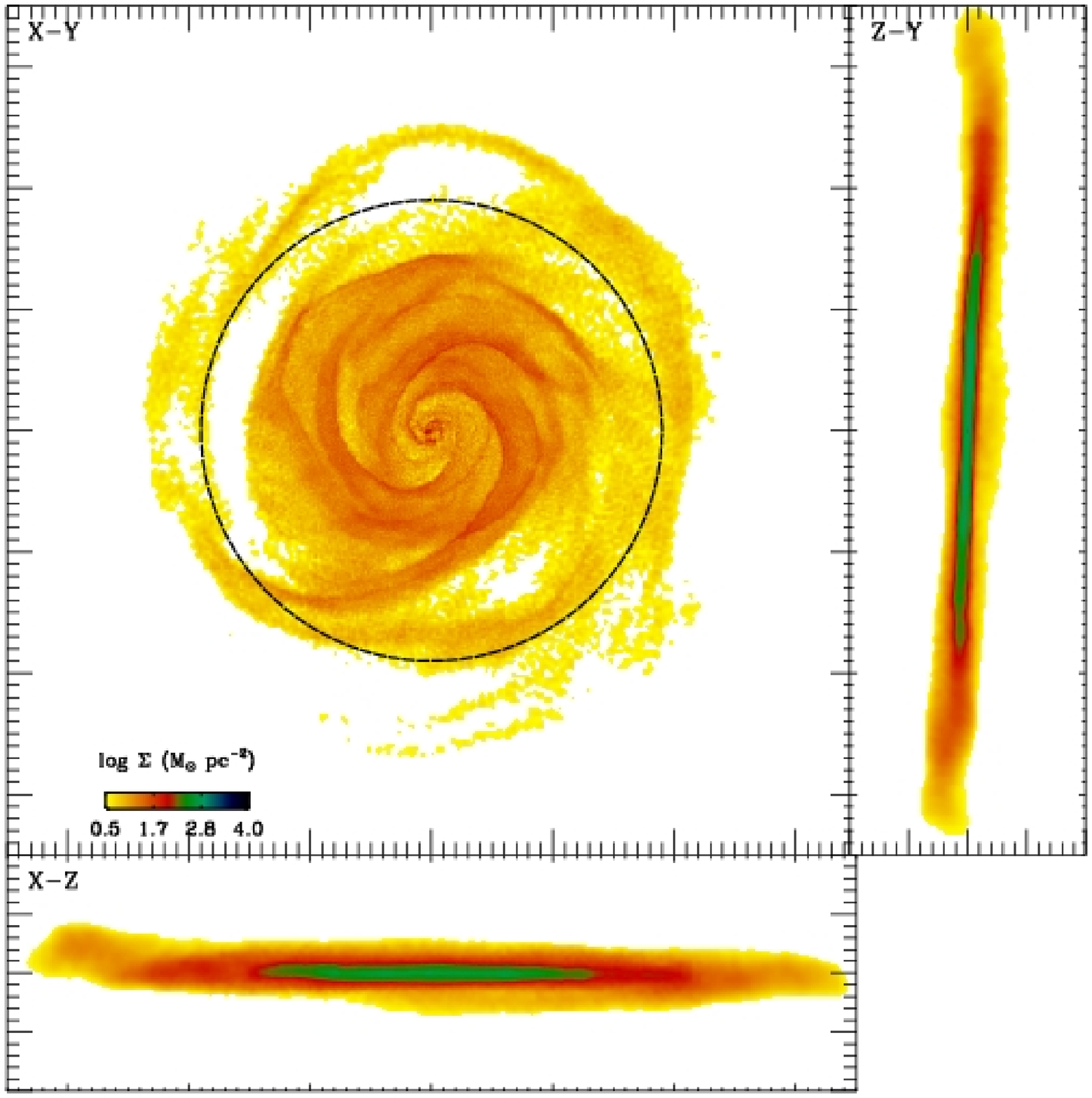}
\caption{Three different projections (X--Y, Z--Y, and X--Z) of the K--band surface brightness (left; again assuming a constant $M/L$ ratio of 2 in solar units) and gas surface mass density (right) at $t\approx 3$ Gyr.  Panels are 50 kpc on a side in the X and Y directions, and 20 kpc in the Z direction.  The solid line corresponds to one of the ring structures as roughly the same galactocentric location ($R_{gac} = 19$ kpc) as the MRi.}
\label{fig:gas}
\end{figure*}

\begin{figure*}
\plottwo{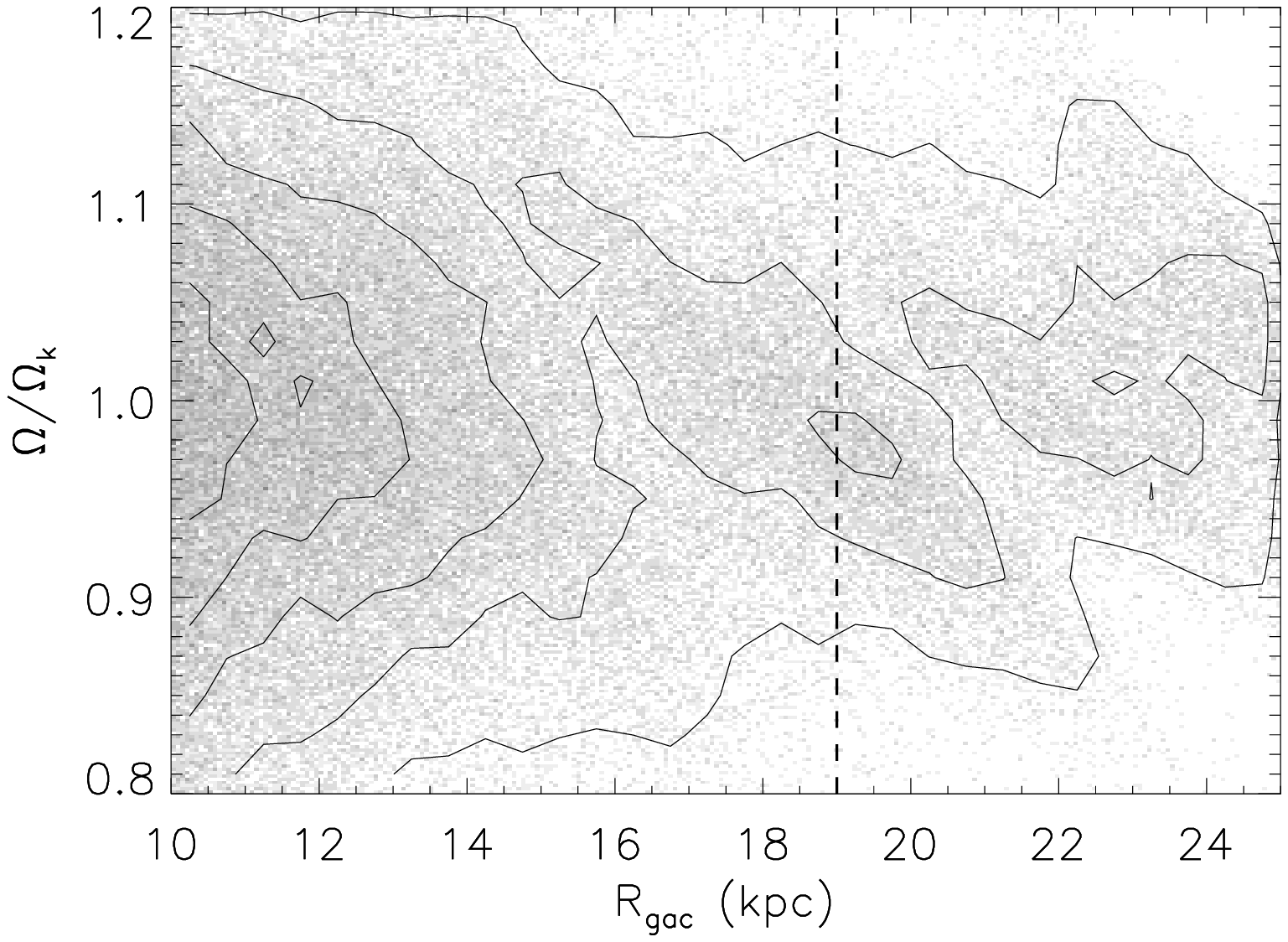}{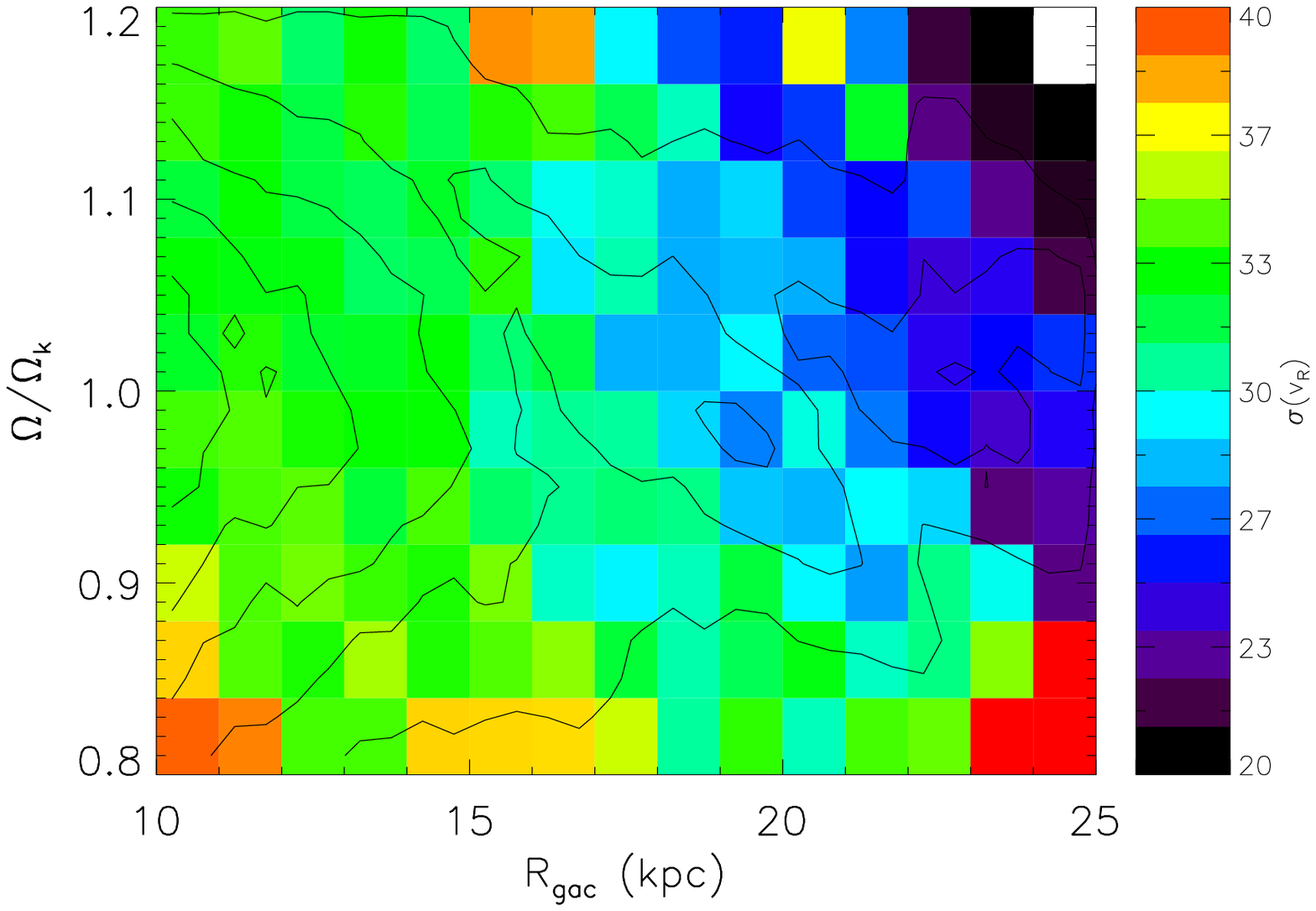}
\caption{Phase space diagram for the stellar particles at $t\approx 3$ Gyr, showing the orbital frequency ($\Omega$) as a fraction of that expected from a circular orbit at that radius ($\Omega_k$) as a function of galactocentric radius ($R_{\rm gac}$).  The left panel shows the density of points in gray-scale, with linear contours overlaid to guide the eye and a dashed line at the location indicated in Figure~\ref{fig:gas}.  The right hand panel is color coded by the radial velocity dispersion ($\sigma(v_R)$) in each cell, with the same contours overlaid.}
\label{fig:phase}
\end{figure*}

The progenitor galaxy models were constructed following \cite{springeldimatteo2005b}, to which we refer the reader for details.  The primary galaxy (total mass of $M_{200} = 10^{12}$\massunitsend) is analogous to the MW with a baronic mass fraction of $m_b=0.05$.  It was realized with $10^6$ halo particles in a \citet{hernquist1990} profile with a concentration of $c=9$ as motivated by cosmological simulations \citep{bullock2001},  and $4\times10^5$ stellar disk (80\% of the baryonic mass) and $2\times10^5$ gas particles (20\% of the baryonic mass).  The satellite galaxy had a total mass of $M_{200}=5\times10^{10}$\massunits ($M_{PG}/M_{SG} \sim 20$) and an identical baryonic mass fraction.  It was realized with $10^5$ halo particles in a \citet{hernquist1990} profile with a concentration of $c=18$, again motivated by cosmological simulations, and $2\times10^4$ stellar disk (50\% of the baryonic mass) and $4\times10^4$ gas particles (50\% of the baryonic mass).  They were placed on a parabolic encounter with $R_p=5$\lengthunits perigalactic radius ($\sim 1$ scale length), consistent with the results of cosmological simulations \citep{benson2005,khochfar2006}.

\section{Discussion}
\label{sec:discuss}

The interaction is summarized in Figures~\ref{fig:allstars} and \ref{fig:gas}.  After the initial close passage, resonances between the orbital frequency of the satellite and primary galaxy disk particles (``stars") excite coplanar tidal arms \citep{toomre1972}.  These features then wrap around the primary galaxy as it continues to revolve, forming a set of concentric ring--like features.  Star formation induced by the merger is concentrated in the nucleus of the primary galaxy \citep[see e.g.,][]{hernquist1989,mihos1994,mihos1996,hernquist1995}, making these rings primarily disk stars.  They are furthermore dynamically cold and kinematically distinct from typical disk stars (see Figure~\ref{fig:phase}), and slowly disperse on a timescale of $\sim 2-4$ Gyr (or several rotation periods) owing to phase mixing of the collisionless stellar particles \citep{binney1987}. Because these rings are generated via gravitational interactions, they are largely insensitive to the gas content or structural parameters of the satellite galaxy.  And because the flyby is effectively an impulse interaction, they will still be formed even if the satellite is disrupted during the encounter.

\section{Relevance to the Monoceros Ring}
\label{sec:monoceros}

The ring features that are produced by the flyby interaction are similar in several ways to the MRi first identified by \citet{newberg2002}.  They provide a good match to the kinematics and location of the ring, and are roughly consistent with metallically measurements of MRi stars.  And, cosmological simulations indicate that such interactions are common for MW--sized halos over the timescale for ring formation \citep[i.e., within the past $\sim 4$ Gyr;][]{stewart2007}.  Therefore, while we do not claim to have modeled the MRi in detail, based on these similarities we propose a flyby interaction as a possible formation scenario.

The MRi forms a coherent structure over $\sim 100^\circ$ in galactic longitude from approximately 15--20 kpc from the galactic center \citep{newberg2002,ibata2003,rochapinto2003,martin2004a,conn2005a,conn2007}.   Kinematically, it is distinguished from MW disk stars by its low radial velocity dispersion ($\sim 25$ km s$^{-1}$); the MRi is dynamically cold \citep{crane2003,yanny2003,conn2005b,martin2005,martin2006}.  Proper motion measurements\footnote{\citet{crane2003} measure an overall circular velocity of $\sim 100$ km s$^{-1}$.  However, we note \citep[as in][]{penarrubia2005} that the accuracy of such measurements in physical units is limited to $\Delta v_{\rm perp} \approx 4.74 R_\odot \Delta \mu$ where $R_\odot \approx 8$ kpc is the galactocentric radius of the sun and $\Delta \mu \sim 3-4$ mas yr$^{-1}$ is the typical proper motion measurement error.  We have therefore chosen to interpret these measurements as being generally consistent with low--eccentricity orbits.}  also suggest that the ring stars are in low--eccentricity orbits \citep[e.g.,][]{crane2003,yanny2003}.  We find that our simulations produce a ring in the right galactocentric radial range that is approximately circularly supported -- $0.9 \lsim \Omega/\Omega_k \lsim 1.1$, where $\Omega$ is the orbital frequency and $\Omega_k$ is that expected from circular motion -- with a similarly low radial velocity dispersion (see Figure~\ref{fig:phase}).  This ring also extends $-4 \lsim z \lsim 4$ kpc above and below the galactic plane, which is consistent with detections such as those of \citet{conn2007}.  Although it is unclear what fraction of the ring would be identified in observations, it contains $\lsim 1\%$ ($\lsim 5\times10^8 M_\odot$) of the total stellar mass of the disk, which is consistent with the mass estimates of \citet{yanny2003}.

A disk origin for MRi stars is also broadly consistent with their observed stellar population.  Several spectroscopic studies \citep[e.g.,][]{yanny2003} have found mean metallicities of $[{\rm Fe/H}] = -1.6\pm 0.3$, while \citet{crane2003} find a higher mean metallicity of $[{\rm Fe/H}] \approx -0.4$ using a different tracer population.  This suggests a primarily metal--poor stellar population with a spread in metallicities, and possibly multiple epochs of star formation.  Our modeling indicates that the MRi may have formed from outer disk stars moved outwards by tidal interactions with the satellite galaxy, which is consistent with this metal--poor stellar population \citep{luck2006,yong2006}.  Furthermore, the tidal interaction moves a significant supply of cold gas into the same ring structures (see Figure~\ref{fig:gas}), which provides the raw material for subsequent epochs of star formation.  While the density--dependent prescription in our simulations does not accurately capture some modes of star formation owing to the limitations of the SPH method, prescriptions that include an approximate treatment of shocks \citep[e.g.,][]{barnes2004} may show these multiple star formation episodes.

\begin{figure}
\plotone{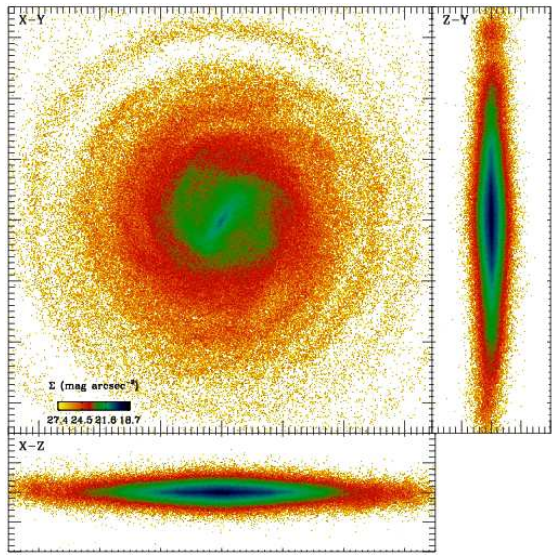}
\caption{Three projections (X--Y, Z--Y, and X--Z) of the K--band surface brightness (again assuming a constant $M/L$ ratio of 2 in solar units) at the same simulation time as Figures~\ref{fig:gas} and \ref{fig:phase} for a $M_{SG}/M_{PG} \approx 1/100$ interaction with identical orbital parameters.}
\label{fig:alt}
\end{figure}

While ring features in our simulations capture many of the properties of the MRi, there are some observations which potentially conflict with our proposal.  In particular, the satellite in our simulations is roughly the same mass as the LMC, and by $\sim 2$ Gyr after the interaction would be at a galactocentric distance of $\sim 250$ kpc.  Such a massive object is well within the detection threshold of current surveys \citep[e.g.,][]{willman2002,koposov2007}; Leo I -- a far lower mass MW companion -- was detected at comparable distance \citep{caputo1999,bellazzini2004,mateo2007}.  However, dynamically cold ring structures are produced generically in flyby encounters, and similar features are present in both a lower--inclination interaction ($i=10^\circ$) and for a lower mass satellite galaxy ($M_{SG}/M_{PG} \sim 1/100$; Figure~\ref{fig:alt}).  It is thus possible that the satellite galaxy that produced the MRi is either (1) hidden from view, at low galactic latitude and/or on the other side of the galaxy, or (2) it is lower mass with $1/100 \lsim M_{SG}/M_{PG} \lsim 1/20$.  Therefore, it is not unreasonable to speculate that Leo I, which had at least one passage through the MW disk within the past 2--4 Gyr \citep{sohn2007}, could have excited the MRi we see today.  Moreover, it is likely that this interaction stripped a significant fraction of its mass\footnote{In our simulations, both the 1/20 and 1/100 encounters stripped $\sim50\%$ of the satellite mass.  While the efficiency of this process is sensitive to both the initial structure of the satellite and the orbital parameters of the interaction, this suggests that a a passage through the disk is likely to remove a significant fraction of the satellite's total mass.}, making current dynamical estimates \citep[$M_{\rm Leo\,I}/M_{MW}\sim 1/1000$;][]{mateo2007} a lower limit on its mass at the time of the interaction.  Furthermore, simulations indicate that there are likely $\gsim$ several sub--haloes in the MW halo with $1/100\lsim M_{SH}/M_{PG}\lsim 1/20$ on orbits that take then through the disk \citep{moore1999,giocoli2007,kazantzidis2007}.  Since galaxy formation may be inefficient in low mass haloes \cite[e.g.,][]{haiman1997,barkana1999,kravtsov2004}, they may be significantly underluminous and therefore difficult to identify.  

Finally, there have been suggestions in the literature \citep{crane2003,frinchaboy2004} that some globular clusters (GCs) may be associated with the MRi.  This is also potentially in conflict with our modeling.  However, we note that their physical association with the MRi is somewhat speculative.  Alternatively, there are $\sim$several nearly coplanar GCs at roughly the same radius \citep{harris1996} as progenitor MRi stars in our simulations that they could have also been moved into the rings by the interaction.

\section{Conclusion}

We present a scenario for the production of dynamically cold rings around a disk galaxy such as the MW via a prograde flyby encounter with a satellite galaxy.  Tidal arms excited during close passage coalesce and wrap around the disk of the primary galaxy.  These kinds of interactions are more cosmologically likely than the nearly circular orbits presented by \citet{martin2005} and \citet{penarrubia2005, penarrubia2007}, while dynamical friction is insufficient to circularize the orbit of such a low mass companion.  Our modeling shows rings with similar spatial distribution and kinematics to the MRi.  The disk origin of MRi stars is furthermore broadly consistent with observed stellar populations.   Therefore, we find that a flyby encounter represents a more cosmologically appealing scenario for the production of the MRi and other dynamically cold rings around the MW.

\acknowledgements

Thanks to the anonymous referee for helpful comments, and to M. Ashby, J. Bullock, S. Dutta, P.. F. Hopkins, C. Hayward, and S. Bush for valuable discussions.  This work was supported in part by a grant from the W. M. Keck foundation.

\bibliographystyle{apj}
\bibliography{../../rings}

\end{document}